\def\etal{{\it et al.}}%
\overfullrule 0 pt
\def\water#1{$\rm (H_2O)_{#1}$}%
\def\Dwater#1{$\rm (D_2O)_{#1}$}%
\def\AB{{\it ab initio}}%
\newcount\refn
\refn=0
\def\jcp{J.~Chem.\ Phys.}
\def\pra{Phys.~Rev.~A}
\def\jpc{J.~Phys.~Chem.}
\def\jpca{J.~Phys.~Chem.~A}
\def\molphys{Molec.\ Phys.}

\def\cpl{Chem.\ Phys.\ Lett.}

\def\prb{Phys.\ Rev.~B}

\def\faraday{J. Chem. Soc., Faraday II}

\def\jacs{J.~Amer.\ Chem.\ Soc.}
\def\ijqc{Int.\ J.~Quantum\ Chem.}

\newcount\endnoteno \endnoteno=0
\newbox\endnotes \newcount\endnoteno \endnoteno=0
\def\bookref#1{%
\global\advance\endnoteno by1%
\global\setbox\endnotes=\vbox{\unvbox\endnotes\vskip2pt%
\refno{\the\endnoteno.} #1}}%
\def\endnote#1#2#3#4#5{%
\global\advance\endnoteno by1%
\global\setbox\endnotes=\vbox{\unvbox\endnotes\vskip2pt%
\refno{\the\endnoteno.}#1, #2 {\bf #3}, #4 (#5).}}%
\def\edref#1{\xdef#1{\the\endnoteno}}%
\def\rf#1{[#1]}%
\def\lastref{\rf{\the\endnoteno}}%
\outer\def\references{%
\unvbox\endnotes}%
\newcount\citeno \newcount\check%
\def\cite#1{\ifnum #1 =1 \global\advance\citeno by1[\the\citeno]%
\else\check=\the\citeno\advance\check by1\global\advance\citeno by #1%
[\the\check-\the\citeno]\fi}%

\input plncs.cmm
%
\newdimen\refindent
\setbox0=\hbox{80.\enspace}\refindent=\wd0\relax
\contribution{Rearrangements of Water Dimer and Hexamer}%
\author{David J.~Wales}%
\address{University Chemical Laboratories, Lensfield Road, Cambridge, CB2 1EW, United Kingdom}%
\abstract{Rearrangement mechanisms of the water dimer and the cage form of the water hexamer
are examined theoretically with particular reference to tunneling splittings and
spectroscopy. The three lowest barrier rearrangements of the water dimer are characterized
by {\it ab initio\/} methods and compared with the results of previous constrained calculations.
The acceptor-tunneling pathway does not proceed via a direct rotation around
the $C_2$ axis of the acceptor, but rather via relatively asynchronous rotation
of the donor about the hydrogen bond and an associated `wag' of the acceptor.
Rearrangements between different cage isomers of the water hexamer are studied for two empirical
potentials. The experimentally observed triplet splittings may be the result of flip
and bifurcation rearrangements of the two single-donor, single-acceptor monomers.
Two-dimensional quantum calculations of the nuclear dynamics suggest that delocalization over
more than one cage isomer may occur, especially in excited states.
\bigskip
{This paper was written in October 1996 for {\it Theory of Atomic and Molecular Clusters} edited 
by Julius Jellinek, which has unfortunately still failed to appear.}}%

\titlea{1}{Introduction}%
Water clusters have proved to be attractive systems for study to both theory and experiment, particularly
in the last two decades. This popularity may be ascribed in part to the role of water as an almost
universal solvent in chemistry and biochemistry, but also to the importance of developing a
fundamental understanding of intermolecular forces \cite4. 
\endnote{D.~D.~Nelson, G.~T.~Fraser and W.~Klemperer}{Science}{238}{1670}{1987}%
\bookref{G.~C.~Maitland, M.~Rigby, E.~B.~Smith and W.~A.~Wakeham, {\it%
Intermolecular Forces\/} (Clarendon Press; Oxford, 1981).}%
\endnote{R.~C.~Cohen and R.~J.~Saykally}{Annu.~Rev.~Phys.~Chem.}{42}{369}{1991}%
\endnote{J.~M.~Hutson}{Annu.~Rev.~Phys.~Chem.}{41}{123}{1990}%
A recent flurry of activity has been sparked by the advent of far-infrared 
vibration-rotation tunneling (FIR-VRT) spectroscopy \cite4,
where resolutions of up to 1$\,$MHz have been achieved.
\endnote{R.~C.~Cohen and R.~J.~Saykally}{\jpc}{94}{7991}{1990}\edref\FIR
\endnote{N.~Pugliano and R.~J.~Saykally}{\jcp}{96}{1832}{1992}\edref\PandS
\endnote{R.~J.~Saykally and G.~A.~Blake}{Science}{259}{1570}{1993}\edref\SandB
\endnote{K.~Liu, J.~D.~Cruzan and R.~J.~Saykally}{Science}{271}{929}{1996}\edref\Sreview
These results provide a new challenge to theory because the tunneling splittings that can
now be resolved necessitate a global view of the potential energy surface (PES) if they
are to be explained. The large amplitude motions which are typical of such weakly bound
Van der Waals complexes sample regions of the PES that are far removed from the bottom
of potential wells, and provide new information about the nature of the intermolecular forces.
Furthermore, to explain or predict the tunneling splittings theory must characterize not only
minima but transition states and rearrangement mechanisms.

The interplay between theory and experiment is perhaps best illustrated by our recently
improved understanding of the dynamics exhibited by water trimer, which began with the
FIR-VRT experiment of Pugliano and Saykally \cite1.
\endnote{N.~Pugliano and R.~J.~Saykally}{Science}{257}{1937}{1992}\edref\PnS
These authors reported spectra for \Dwater{3}\ characteristic of an oblate symmetric rotor, with each
line split into a regularly spaced quartet and a spacing of roughly $6\,$MHz, i.e.~$2\times10^{-4}\,$cm$^{-1}$.
Accurate values for the vibrationally averaged rotational constants revealed a large negative inertial defect,
indicative of extensive out-of-plane motion of the non-hydrogen-bonded hydrogens.
However, a cyclic, asymmetric global minimum structure for the trimer was established some years
ago in \AB\ calculations \cite1, in agreement with earlier experimental results \cite1.
\endnote{J.~Del Bene and J.~A.~Pople}{\jcp}{58}{3605}{1973}%
\endnote{M.~F.~Vernon, D.~J.~Krajnovich, H.~S.~Kwok, J.~M.~Lisy, Y.~R.~Shen and Y.~T.~Lee}{\jcp}{77}{47}{1982}%

The oblate symmetric top spectrum of the trimer has now been explained by vibrational averaging over large amplitude torsional
motions of the free hydrogens on the timescale of the FIR-VRT experiment.
These large amplitude motions are associated with a facile `flip' rearrangement where a free hydrogen
moves from one side of the plane defined by the three oxygen atoms to the other.
This mechanism was probably first
characterized for an empirical potential by Owicki \etal\ \cite1.
\endnote{Owicki, J.~C.; Shipman, L.~L.; Scheraga, H.~A.}{\jpc}{79}{1794}{1975}\edref\owi
An \AB\ pathway has been presented by Wales \cite1\ and the corresponding transition state was also
characterized by Fowler and Schaefer \cite1.
\endnote{D.~J.~Wales}{\jacs}{115}{11180}{1993}\edref\DJWtrimer
\endnote{J.~E.~Fowler and H.~F.~Schaefer}{\jacs}{117}{446}{1995}\edref\FandS
Wales identified two other degenerate rearrangement mechanisms for the trimer
with rather larger barriers than the flip, and suggested that one of them (christened the `donor' or
`bifurcation' pathway) might be responsible for the quartet splittings observed experimentally \rf\DJWtrimer.
In the associated transition state one monomer acts as a double donor to a neighbour which acts as
a double acceptor in a configuration similar to that of the `donor tunneling' transition state
in the water dimer, discussed in \S 4.

Subsequent experiments \cite2\ assigned new transitions for \water{3}\ and \Dwater{3}\ which revealed
rigorously symmetric rotor structure, in contrast to the strongly perturbed band 
that was initially investigated by Pugliano and Saykally \rf\PnS.
\endnote{K.~Liu, J.~G.~Loeser, M.~J.~Elrod, B.~C.~Host, J.~A.~Rzepiela, N.~Pugliano and
   R.~J.~Saykally}{\jacs}{116}{3507}{1994}\edref\liu
\endnote{S.~Suzuki and G.~A.~Blake}{\cpl}{229}{499}{1994}\edref\Suzuki
All the transitions reported by Liu \etal\ \rf\liu\ and Suzuki and Blake \rf\Suzuki\ show
a regular quartet splitting of every rovibrational transition.
Liu \etal\ reported that their new spectra were consistent with the $G(48)$ group suggested by the
mechanisms discovered by Wales \rf\DJWtrimer. Walsh and Wales have recently studied the bifurcation
mechanism in more detail and discovered that six slightly different alternatives exist, depending
upon the level of the calculation employed \cite1.
\endnote{T.~R.~Walsh and D.~J.~Wales}{J.~Chem.~Soc., Faraday Trans.}{92}{2505}{1996}\edref\WWtri

Numerous other treatments of the water trimer have focused on the breakdown of the total energy
into many-body contributions and the effects of electron correlation \cite5, 
the development of torsional potential energy surfaces \cite3, the
calculation of torsional energy levels using the discrete variable representation (DVR) \cite4\ 
and quantum simulations of the nuclear dynamics using the diffusion Monte
Carlo (DMC) method \cite2.
\endnote{J.~G.~C.~M.~van~Duijneveldt-van~de~Rijdt and F.~B.~van~Duijneveldt}{\jcp}{97}{5019}{1982}%
\endnote{G.~Chalasi\' nski, M.~M.~Szcz\c e\'sniak, P.~Cieplak and S.~Scheiner}{\jcp}{94}{2873}{1991}\edref\CSCS
\endnote{S.~S.~Xantheas and T.~H.~Dunning}{\jcp}{98}{8037}{1993}\edref\XandDa
\endnote{S.~S.~Xantheas and T.~H.~Dunning}{\jcp}{99}{8774}{1993}\edref\XandDb
\endnote{S.~S.~Xantheas}{\jcp}{100}{7523}{1994}\edref\XMB
\endnote{van~Duijneveldt-van~de~Rijdt, J.~G.~C.~M.; van~Duijneveldt, F.~B.}{\cpl}{237}{560-567}{1995}\edref\vdj
\endnote{W.~Klopper, M.~Sch\"utz, H.~P.~L\"uthi and Leutwyler, S.}{\jcp}{103}{1085}{1995}\edref\surf
\endnote{T.~B\"urgi, S.~Graf, S.~Leutwyler and W.~Klopper}{\jcp}{103}{1077}{1995}\edref\huckel
\endnote{M.~Sch\"utz, T.~B\"urgi, S.~Leutwyler and H.~B.~B\"urgi}{\jcp}{99}{5228}{1993}\edref\swis
\endnote{M.~Sch\"utz, T.~B\"urgi, S.~Leutwyler and H.~B.~B\"urgi}{\jcp}{100}{1780}{1994}\edref\error
\endnote{W.~Klopper and M.~Sch\"utz}{\cpl}{237}{536}{1995}\edref\klop
\endnote{D.~Sabo, Z.~Ba\v ci\'c, T.~B\"urgi and S.~Leutwyler}{\cpl}{244}{283}{1995}\edref\Bacic
\endnote{J.~K.~Gregory and D.~C.~Clary}{\jcp}{102}{7817}{1995}\edref\GandC
\endnote{J.~K.~Gregory and D.~C.~Clary}{\jcp}{103}{8924}{1995}\edref\GandC

In the present contribution we report new results for the water dimer and water hexamer, which lie at the
two limits of the size range for which high resolution data have been obtained. Unconstrained pathways
do not seem to have been described before for the three rearrangements of the water dimer which lead to 
observable splittings or shifts in the pattern of rovibrational energy levels. In particular, we find
that the acceptor tunneling mechanism is not simply a rotation about the local $C_2$ axis of the acceptor
monomer. For water hexamer we employ two empirical intermolecular potentials to
examine the cage isomers and show that these structures are connected by facile, non-degenerate
single flips of the two single-donor, single-acceptor monomers. Two-dimensional DVR calculations are 
performed to investigate these motions.
First we give a brief overview of the effective molecular symmetry group which must be employed
to classify the energy levels of floppy molecules.

\titlea{2}{The Effective Molecular Symmetry Group}%
The observation of quantum tunneling effects in high resolution spectra provides indirect information
about the corresponding rearrangement mechanisms. To make use of this information we need 
to employ group theory to classify the energy levels of such systems. Point groups prove
satisfactory when the molecule under consideration is rigid on the appropriate experimental timescale.
However, molecular potential energy surfaces generally contain local minima corresponding to 
permutational isomers of any given stationary point. We will adopt the nomenclature of Bone \etal\ \cite1\
where a \lq structure' is understood to mean a particular molecular geometry and a \lq version' is a
particular labelled permutational isomer of a given structure.
\endnote{R.~G.~A.~Bone, T.~W.~Rowlands, N.~C.~Handy and A.~J.~Stone}{\molphys}{72}{33-73}{1991}\edref\Bone
Minima which are directly connected by a given rearrangement are said to be \lq adjacent'.

Tunneling splittings may be observed when rovibronic wavefunctions localized in the potential wells
corresponding to different versions interfere with each other. For example, the ammonia molecule
displays doublet splittings because pairs of permutational isomers are interconverted by the
inversion mechanism in which the molecule passes through a planar transition state \cite2.
\endnote{D.~M.~Dennison and J.~D.~Hardy}{Phys Rev.}{39}{938-947}{1932}%
\bookref{R.~P.~Bell, {\it The Tunnel Effect in Chemistry\/} (Chapman and Hall; New York, 1980).}\edref\Bell
If the barrier corresponding to a certain rearrangement mechanism is sufficiently low, the path length
sufficiently short, and the associated effective mass sufficiently small then tunneling may occur.
Hence tunneling splittings can usually be associated with a low energy transition state corresponding to 
a degenerate rearrangement mechanism \cite1\ which links different versions of the same structure, as opposed
to different structures where the energy levels would not be in resonance.
\endnote{R.~E.~Leone and P.~v.~R.~Schleyer}{Angew.~Chem.~Int.~Ed.~Engl.}{9}{860}{1970}%
Degenerate rearrangements can be either symmetric or asymmetric, depending upon whether the two
sides of the corresponding path are related by symmetry \cite1.
\endnote{J.~G.~Nourse}{\jacs}{102}{4883}{1980}\edref\nourse
We follow Murrell and Laidler's definition of a transition state as a stationary point with a single
negative Hessian eigenvalue \cite1.
\endnote{J.~N.~Murrell and K.~J.~Laidler}{\faraday}{64}{371}{1968}\edref\MandL

The energy levels of non-rigid molecules can be classified using the Complete Nuclear Permutation-Inversion 
(CNPI) group which is the direct product of the group containing all possible permutations
of identical nuclei and the inversion group. The latter group contains only the identity operation, $E$,
and the operation of inversion
of all particle coordinates through the space-fixed origin, $E^*$, which commutes with all the
permutations. The CNPI group is a true symmetry
group of the full molecular Hamiltonian in the absence of external fields, and its elements are generally referred to as
permutation-inversion operations. However, the order of this group increases factorially for systems
containing increasing numbers of atoms of the same element, and rapidly becomes difficult to use.
Fortunately, Longuet-Higgins showed that it not necessary to consider the whole of the CNPI group, but
rather only the permutation-inversions which correspond to tunneling splittings that are resolvable for
a given experiment \cite1.
\endnote{H.~C.~Longuet-Higgins}{\molphys}{6}{445}{1963}%
The corresponding rearrangement mechanisms are said to be `feasible'.
The resulting subgroup of the CNPI group is known as the effective molecular symmetry (MS) group.
For rigid molecules the appropriate MS group is isomorphic to the usual rigid molecule point group \cite4.
\endnote{J.~T.~Hougen}{\jcp}{37}{1433}{1962}%
\endnote{J.~T.~Hougen}{\jcp}{39}{358}{1962}%
\endnote{J.~T.~Hougen}{\jpc}{90}{562}{1986}%
\bookref{P.~R.~Bunker, {\it Molecular Symmetry and Spectroscopy\/} (Academic Press; New York, 1970).}\edref\bun
The corresponding permutation-inversions simply correspond to overall rotation of the system, and are
therefore always feasible.

If there exists a non-trivial feasible rearrangement with a finite barrier then the MS group is
enlarged and the wavefunctions which transform according to irreducible representations of this
group are linear combinations of the functions localized in each well. In general, a given mechanism
will not link all the possible versions of a given structure but rather the versions will be
partitioned into a number of closed sets with equivalent reaction graphs for each set.
If we consider a representative set of versions all of which can be interconverted by repeated application
of the feasible rearrangement, then the corresponding wavefunction must be a linear combination
of the localized functions from members of this set. The delocalized wavefunctions can be found
by solving a secular problem, just as in the linear combination of molecular orbitals approach
to electronic structure.
If the wavefunctions decay rapidly in the classically forbidden regions of the PES between minima
then it may be sufficient to consider only nearest-neighbour interactions and the resulting
splitting pattern is then determined largely by symmetry.
The connectivity of the reaction graph, the associated MS group and the splitting pattern can all
be found automatically by a computer program once a minimal set of generator permutation-inversions
is known \cite1.
\endnote{D.~J.~Wales}{\jacs}{115}{11191}{1993}\edref\DJWpert

\titlea{3}{Geometry Optimizations and Potentials}%
The geometry optimizations and calculations of
rearrangement pathways described in the following sections were all
performed by eigenvector-following \cite1.
\bookref{C.~J.~Cerjan and W.~H.~Miller, \jcp\ {\bf75}, 2800 (1981);
 J.~Simons, P.~J\o rgenson, H.~Taylor and J.~Ozment, \jpc\ {\bf 87}, 2745 (1983);
 D.~O'Neal, H.~Taylor and J.~Simons, \jpc\ {\bf88}, 1510 (1984);
 A.~Banerjee, N.~Adams, J.~Simons and R.~Shepard, \jpc\ {\bf89}, 52 (1985);
 J.~Baker, J.~Comput.~Chem.~{\bf7}, 385 (1986);
 J.~Baker, J.~Comput.~Chem.~{\bf8}, 563 (1987).}\edref\CM
Details of the precise implementation have been given elsewhere \cite2.
\endnote{D.~J.~Wales}{\jcp}{101}{3750}{1994}\edref\fiftyfive
\endnote{D.~J.~Wales and T.~R.~Walsh}{\jcp}{105}{6957}{1996}\edref\WWpent
Analytic first and second derivatives of the energy were calculated at every step.
In the \AB\ calculations these derivatives were all generated by the CADPAC program \cite1,
and Cartesian coordinates were used throughout. 
\bookref{R.~D.~Amos and J.~E.~Rice,~{\it CADPAC: the Cambridge Analytic
Derivatives Package, Issue 4.0\/}; Cambridge, 1987.}\edref\cadpac
Pathways were calculated by taking small displacements of $0.03\,a_0$ away from a transition state
both parallel and antiparallel to the transition vector, and then employing eigenvector-following 
energy minimization to find the associated minimum. The pathways obtained by this procedure
have been compared to steepest-descent paths and pathways that incorporate a kinetic metric \cite1\
in previous work---the mechanism is generally found to be represented correctly \rf\WWtri.
\endnote{A.~Banerjee and N.~P.~Adams}{Int.~J.~Quant.~Chem.}{43}{855}{1992}%

Calculations employing rigid body intermolecular potentials were performed using the ORIENT3 program \cite3,
which contains the same optimization package adapted for centre-of-mass/orientational coordinates.
\endnote{P.~L.~A.~Popelier, A.~J.~Stone and D.~J.~Wales}{J.~Chem.~Soc., Faraday Discuss.}{97}{243}{1994}%
\endnote{D.~J.~Wales, P.~L.~A.~Popelier and A.~J.~Stone}{\jcp}{102}{5556}{1995}%
\endnote{D.~J.~Wales, A.~J.~Stone and P.~L.~A.~Popelier}{\cpl}{240}{89}{1995}%
This program can treat intermolecular potentials based upon Stone's
distributed multipoles \cite2\ and distributed polarizabilities \cite2; simpler
models based upon point charges and Lennard-Jones interactions fall within this framework.
\endnote{A.~J.~Stone}{\cpl}{83}{233}{1981}%
\endnote{A.~J.~Stone and M.~Alderton}{\molphys}{56}{1047}{1985}\edref\DMA%
\endnote{A.~J.~Stone}{\molphys}{56}{1065}{1985}%
\endnote{C.~R.~Le Sueur and A.~J.~Stone}{\molphys}{78}{1267}{1993}%
Calculations for the water hexamer in \S 5 were performed using the relatively sophisticated
ASP-W2 potential of Millot and Stone \cite1\ (somewhat modified from the published version)
and the much simpler but widely-used TIP4P form \cite2.
\endnote{C.~Millot and A.~J.~Stone}{\molphys}{77}{439}{1992}\edref\ASP
\endnote{W.~L.~Jorgensen}{\jacs}{103}{335}{1981}%
\endnote{W.~L.~Jorgensen, J.~Chandraesekhar,  J.~W.~Madura, R.~W.~Impey and M.~L.~Klein}%
 {\jcp}{79}{926}{1983}\edref\TIP
\begfig 13.3 cm
\figure{1}{Acceptor-tunneling path for \water{2}\ calculated at the DZP+diff/BLYP level. For
this path $S=4.2\,a_0$, $D=2.8\,a_0$ and $\gamma=1.3$ (see \S 3).}%
\endfig

In the \AB\ Hartree-Fock (HF) calculations for \water{2}\ two basis sets were considered. The smaller
double-$\zeta$ \cite2\ plus polarization (DZP) basis employed polarization functions consisting of
a single set of $p$ functions on each hydrogen atom (exponent 1.0) and a set of
six $d$ functions on each oxygen atom (exponent 0.9) to give a total of 26 basis functions
per monomer.
\endnote{T.~H.~Dunning Jr.}{\jcp}{53}{2823}{1970}%
\endnote{S.~J.~Huzinaga}{\jcp}{47}{1293}{1965}\edref\huz
The larger basis set, denoted DZP+diff, includes the above DZP functions 
with an additional diffuse $s$ function on each
hydrogen atom (exponent 0.0441) and diffuse sets of $s$ and $p$ functions on each oxygen atom (exponents 0.0823 and 0.0651
for $s$ and $p$ respectively) \rf\FandS, to give 32 basis functions per monomer.
Correlation corrections were obtained through both second order M\o ller-Plesset (MP2) theory \cite1\
and density functional theory (DFT).
\endnote{C.~M\o ller and M.~S.~Plesset}{Phys.~Rev.}{46}{618}{1934}\edref\MandP
In the DFT calculations we employed the Becke nonlocal exchange functional \cite1\
and the Lee-Yang-Parr correlation functional \cite1\ (together referred to as BLYP); derivatives of the grid
weights were not included and the core electrons were not frozen.
\endnote{A.~D.~Becke}{\pra}{38}{3098}{1988}%
\endnote{C.~Lee, W.~Yang and R.~G.~Parr}{\prb}{37}{785}{1988}%
Numerical integration of the BLYP functionals was performed using grids between the CADPAC \lq MEDIUM' and
\lq HIGH' options. The \lq MEDIUM' grids were not accurate enough to give the right number of negative
Hessian eigenvalues, whereas the \lq HIGH' grids contained more points than necessary.
CADPAC actually uses different sized grids for different parts of the calculation \rf\cadpac; in the present
work these grids contained 14,386 and 97,008 points after removal of those with densities below the 
preset tolerances.
Calculations were deemed to be converged when the root-mean-square gradient fell below $10^{-6}$ atomic units.
This is sufficient to reduce the six `zero' normal mode frequencies to less than $1\,$cm$^{-1}$ in the
HF and MP2 calculations. Because derivatives of the grid weights were not included, the largest of the
six `zeros' can be as big as $20\,$cm$^{-1}$ for the DFT stationary points.

Three additional parameters are useful in describing the rearrangement mechanisms.
The first is the integrated path length, $S$, which was calculated as a sum over eigenvector-following steps.
The second is the distance between the two minima in nuclear configuration space, $D$.
The third is the moment ratio of displacement \cite1, $\gamma$, which gives a measure of the
cooperativity of the rearrangement:
\endnote{F.~H.~Stillinger and T.~A.~Weber}{Phys.~Rev.~A}{28}{2408}{1983}\edref\SWgamma
$$ \gamma = { N \sum_i \left[{\bf Q}_i(s)-{\bf Q}_i(t)\right]^4  \over
   \left( \sum_i \left[{\bf Q}_i(s)-{\bf Q}_i(t)\right]^2 \right)^2 }, \eqno(1) $$
where ${\bf Q}_i(s)$ is the position vector in Cartesian coordinates
for atom $i$ in minimum $s$, etc., and $N$ is the number of atoms.
If every atom undergoes the same displacement in one Cartesian component then
$\gamma=1$, while if only one atom has one non-zero component then $\gamma=N$, i.e.~18 for \water6.
\begfig 13.3 cm
\figure{2}{Donor-acceptor-interchange tunneling mechanism for \water{2}\ calculated at the
DZP+diff/BLYP level.  For this path $S=5.4\,a_0$, $D=4.4\,a_0$ and $\gamma=1.4$ (see \S 3).}%
\endfig

\titlea{4}{Rearrangements of Water Dimer}%
There have been many experimental investigations of the water dimer
\cite1\ following the original work of Dyke, Mack and Muenter \cite1.
\endnote{G.~T.~Fraser}{Int.~Rev.~Phys.~Chem.}{10}{189}{1991}%
\endnote{T.~R.~Dyke, K.~M.~Mack and J.~S.~Muenter}{\jcp}{66}{498}{1977}\edref\DMM
Most recently, a complete characterization of the tunneling dynamics in a vibrationally excited state
of \Dwater{2}\ has been presented \cite1.
\endnote{N.~Pugliano, J.~D.~Cruzan, J.~G.~Loeser and R.~J.~Saykally}{\jcp}{98}{6600}{1993}\edref\Sdimer
The dimer rovibronic energy levels were first classified in terms of permutation-inversion group theory
by Dyke \cite1, and Coudert and Hougen applied their internal axis approach to the intermolecular
dynamics using an empirical potential \cite3.
\endnote{T.~R.~Dyke}{\jcp}{66}{492}{1977}\edref\Dyke
\endnote{J.~T.~Hougen}{J.~Mol.~Spectr.}{114}{395}{1985}\edref\Hougen
\endnote{L.~H.~Coudert and J.~T.~Hougen}{J.~Mol.~Spectr.}{130}{86}{1988}\edref\CandHa
\endnote{L.~H.~Coudert and J.~T.~Hougen}{J.~Mol.~Spectr.}{139}{259}{1990}\edref\CandHb
Smith \etal\ \cite1\ performed \AB\ calculations and identified three true transition states for
\water{2}; they also performed constrained calculations of the donor-tunneling pathway.
\endnote{B.~J.~Smith, D.~J.~Swanton, J.~A.~Pople, H.~F.~Schaefer and L.~Radom}{\jcp}{92}{1240}{1990}\edref\Pople
In this section we will describe unconstrained pathway calculations for all three rearrangements;
the energetics of the various stationary points at different levels of theory are summarized in
Table 1 and counterpoise-corrected \cite1\ binding energies (including monomer relaxation \cite1)
are given in Table 2.
\endnote{S.~F.~Boys and F.~Bernardi}{\molphys}{19}{553}{1970}\edref\counter
\endnote{S.~S.~Xantheas}{\jcp}{104}{8821}{1996}\edref\bsserelax
\begfig 13.3 cm
\figure{3}{Donor-tunneling path for \water{2}\ calculated at the DZP+diff/BLYP level.
For this path $S=5.8\,a_0$, $D=4.5\,a_0$ and $\gamma=2.2$ (see \S 3).}%
\endfig

There are a total of $2\times2!\times4!/2=48$ distinct versions of the water dimer global minimum
on the PES (we divide by two because the rigid molecule point group has order two \rf\Bone).
However, mechanisms which involve the making and breaking of covalent bonds lie much too 
high in energy to give rise to observable tunneling splittings. The largest possible MS group which
can pertain when covalent bond-breaking is not feasible has order $2\times2!\times(2!)^2=16$, 
where the first factor accounts for the inversion mechanism,
the second factor accounts for permutation of the two oxygen nuclei, and the last term accounts
for the permutation of the two hydrogen (or deuterium) atoms within each monomer. This group may be
denoted $G(16)$ and is isomorphic to the point group $D_{4h}$ \rf\Dyke. Since the equilibrium
geometry has $C_s$ symmetry the maximum number of distinct 
versions that can be interconverted without breaking covalent bonds is $16/2=8$.
\vskip 8 true mm \tabcap{1}{Energies in hartree and point groups at various levels of theory 
for the water dimer global minimum
and the transition states for acceptor tunneling, donor-acceptor interchange and donor tunneling.}%
\centerline{
$$\vbox{\petit
\halign{
   #     \hfil&
  $#$\ \hfil&
  $#$\ \hfil&
  $#$\ \hfil&
  $#$\hfil\cr
   \noalign{\medskip}%
   \noalign{\hrule}%
   \noalign{\vskip1.5true pt}%
   \noalign{\hrule}%
   \noalign{\medskip}%
   & \hfil$DZP/HF$  & \hfil$DZP+diff/HF$  & \hfil$DZP+diff/BLYP$    & \hfil$DZP+diff/MP2$   \cr
   \noalign{\medskip}%
   \noalign{\hrule}%
   \noalign{\vskip1.5true pt}%
   \noalign{\hrule}%
   \noalign{\medskip}%
 minimum & -152.102057 (C_s)    & -152.107971 (C_s)    &  -152.87817 (C_s)    & -152.540770 (C_s) \cr
 acceptor& -152.101274 (C_s)    & -152.107282 (C_s)    &  -152.87701 (C_1)    & -152.539673 (C_1) \cr
 don-acc & -152.100688 (C_i)    & -152.106287 (C_i)    &  -152.87591 (C_i)    & -152.538743 (C_i) \cr
 donor   & -152.099832 (C_{2v}) & -152.105946 (C_{2v}) &  -152.87535 (C_{2v}) & -152.538149 (C_{2v}) \cr
  \noalign{\medskip\hrule\vskip1.5true pt\hrule}%
 }  } $$
}%
\vskip 8 true mm
In the equilibrium $C_s$ geometry one `donor' monomer acts as a single hydrogen-bond donor to the other
`acceptor' molecule \cite1. 
\endnote{J.~A.~Odutola and T.~R.~Dyke}{\jcp}{72}{5062}{1980}%
The largest tunneling splitting is due to a mechanism in which the two hydrogen atoms of the
acceptor monomer are effectively interchanged, for which Smith \etal\ \rf\Pople\ reported a barrier of 206$\,$cm$^{-1}$.
In the present work a planar transition state of $C_s$ symmetry was found for 
both basis sets in the HF calculations, but this changed to an out-of-plane $C_1$ structure when
correlation corrections were applied, in agreement with Smith \etal\ \rf\Pople. 
The pathway is shown in Fig.~1, and corresponds to
a `methylamine-type' process \cite1\ rather than a direct rotation about the local $C_2$ axis of 
the acceptor monomer.
The path is represented by nine snapshots where the first and last
frames are the two minima, the middle frame is the transition state and three additional frames on
each side of the path were selected to best illustrate the mechanism. All the pathways in the present work
were visualized using Mathematica \cite1.
\endnote{M.~Tsuboi, A.~Y.~Hirakawa, T.~Ino, T.~Sasaki and K.~Tamagake}{\jcp}{41}{2721}{1964}%
\bookref{Mathematica 2.0; Wolfram Research Inc., Champaign, IL, 1989.}%

The above mechanism is in agreement with the analysis of Pugliano \etal\ \rf\Sdimer\ for the ground
state acceptor tunneling path. The generator permutation-inversion corresponding to the labelling
scheme of Fig.~1 is (34), and it connects the versions in pairs. Hence each rigid-dimer rovibrational
level is split into two by this process. Experimentally the ground state splitting 
due to acceptor tunneling is \cite1\
$2.47\,$cm$^{-1}$, in good agreement with a five-dimensional treatment of the nuclear dynamics
by Althorpe and Clary \cite1, which yielded a value of $2.34\,$cm$^{-1}$.
The MS group for the rigid dimer labelled according to Fig.~1 contains only the identity, $E$, and the
permutation-inversion (34)*, where hydrogens three and four change places and all
coordinates are inverted 
through the space-fixed origin. The appropriate MS group when the generator (34)
operation is feasible contains the elements $E$, $E$*, (34) and (34)*.
\endnote{E.~Zwart, J.~J.~ter Muelen, W.~L.~Meerts and L.~H.~Coudert}{J.~Mol.~Spectrosc.}{147}{27}{1991}\edref\ZMMC
\endnote{S.~C.~Althorpe and D.~C.~Clary}{\jcp}{101}{3603}{1995}\edref\AandC
\vskip 8 true mm \tabcap{2}{Couterpoise-corrected \rf\counter\ binding energies in millihartree 
at various levels of theory for the water dimer global minimum
and the transition states for acceptor tunneling, donor-acceptor interchange and donor tunneling.}%
\centerline{
$$\vbox{\petit
\halign{
         # \hfil&
 \  \hfil$#$\ \hfil&
 \  \hfil$#$\ \hfil&
 \  \hfil$#$\ \hfil&
 \  \hfil$#$\hfil\cr
   \noalign{\medskip}%
   \noalign{\hrule}%
   \noalign{\vskip1.5true pt}%
   \noalign{\hrule}%
   \noalign{\medskip}%
   & \hfil$DZP/HF$  & \hfil$DZP+diff/HF$  & \hfil$DZP+diff/BLYP$    & \hfil$DZP+diff/MP2$   \cr
   \noalign{\medskip}%
   \noalign{\hrule}%
   \noalign{\vskip1.5true pt}%
   \noalign{\hrule}%
   \noalign{\medskip}%
 minimum & 7.42  & 6.92  & 7.75  & 7.45 \cr
 acceptor& 6.88  & 6.38  & 6.64  & 6.65 \cr
 don-acc & 5.73  & 5.82  & 6.03  & 6.52 \cr
 donor   & 5.13  & 5.07  & 4.79  & 5.28 \cr
  \noalign{\medskip\hrule\vskip1.5true pt\hrule}%
 }  } $$
}%
\vskip 8 true mm
\begfig 9.1 cm
\figure{4}{Energy versus the integrated path length, $S$, for the three degenerate
rearrangement mechanisms of the water dimer described in \S 4.}%
\endfig

The next largest splitting is caused by donor-acceptor interchange. 
For this process Smith \etal\ \rf\Pople\ 
found a cyclic transition state with $C_i$ symmetry and an associated  barrier of $304\,$cm$^{-1}$. 
In this rearrangement, for which the calculated pathway is shown in Fig.~2,
the roles of the donor and acceptor monomers are interchanged. An appropriate  generator for
the labelling scheme of Fig.~2 is (AB)(1423), i.e.~oxygen A is replaced by oxygen B, hydrogen
1 is replaced by hydrogen 4, hydrogen 4 is replaced by hydrogen 2 etc. This generator is
not unique---one could also choose (AB)(14)(23)* and obtain the same MS group. 

If donor-acceptor
interchange is the only feasible mechanism then the versions are connected in sets of four and
the MS group has eight members: class 1 contains $E$, class 2 contains (12)(34), class 3 contains
(AB)(1423) and (AB)(1324), class 4 contains (34)* and (12)*, and class 5 
contains (AB)(14)(23)* and (AB)(13)(24)*.
The splitting pattern in the simplest H\"uckel-type approximation \cite1\ is:
\endnote{D.~J.~Wales}{\jacs}{115}{11191}{1993}\edref\DJWtun
$$ 2\beta_{\rm da} (A_1), \qquad 0 (E), \qquad -2\beta_{\rm da} (B_1), \eqno(2) $$
where $\beta_{\rm da}$ is the donor-acceptor-interchange tunneling matrix element and
we have labelled the levels according to appropriate irreducible representations of the
$G(8)$ group which is a subgroup of $G(16)$ \rf\Dyke.
When this process and 
acceptor tunneling are both feasible the MS group has order 16, i.e.~the largest MS group
possible without breaking covalent bonds, and is isomorphic to $D_{4h}$ \rf\Dyke. 
The versions are then connected in sets of eight and the splitting pattern is:
$$ \eqalignno{
  &\beta_{\rm a}+2\beta_{\rm da} (A_1^+), \qquad 
   \hphantom{-}\beta_{\rm a} (E^+), \qquad 
   \hphantom{-}\beta_{\rm a}-2\beta_{\rm da} (B_1^+),  \cr
  -&\beta_{\rm a}+2\beta_{\rm da} (A_2^-), \qquad
  -\beta_{\rm a} (E^-), \qquad
  -\beta_{\rm a}-2\beta_{\rm da} (B_2^-), &(3) \cr }  $$
where $\beta_{\rm a}$ is the acceptor-tunneling  matrix element.
Experimentally, the tunneling splittings due to donor-acceptor-interchange are about a
factor of five smaller than those associated with acceptor tunneling. 

The third process which is generally presumed to have an observable effect on the energy level
diagram is donor tunneling, but this can only lead to energy level shifts rather than further
splittings because $G(16)$ is the largest MS group possible without breaking covalent bonds. 
Smith \etal\ \rf\Pople\ calculated a barrier of  $658\,$cm$^{-1}$ for this mechanism, and the
corresponding pathway found in the present work is shown in Fig.~3.
An appropriate generator permutation-inversion for the labelling scheme of Fig.~3 is (12)(34),
and on its own this process would simply lead to doublet splittings with versions linked in
pairs and an MS group of order 4.  However, when combined with the other two mechanisms the
eigenvalues of the corresponding secular problem (assuming a H\"uckel-type approximation \rf\DJWtun) are:
$$ \eqalignno{
   &\beta_{\rm a}+2\beta_{\rm da}+\beta_{\rm d} (A_1^+), \qquad 
    \hphantom{-}\beta_{\rm a}-\beta_{\rm d} (E^+), \qquad 
    \hphantom{-}\beta_{\rm a}-2\beta_{\rm da}+\beta_{\rm d} (B_1^+), \  \cr
  -&\beta_{\rm a}+2\beta_{\rm da}+\beta_{\rm d} (A_2^-), \qquad
   -\beta_{\rm a}-\beta_{\rm d} (E^-), \qquad
   -\beta_{\rm a}-2\beta_{\rm da}+\beta_{\rm d} (B_2^-). &(4) \cr } $$
This pattern is in complete agreement with that obtained by Althorpe and Clary \rf\AandC\ and
with the model calculations of Coudert and Hougen [\Hougen-\CandHb]. 
Since the tunneling levels are no longer in plus-minus pairs we conclude that the
presence of donor tunneling introduces odd-membered rings into the reaction graph \cite1.
\endnote{C.~A.~Coulson and S.~Rushbrooke}{Proc.~Camb.~Phil.~Soc.}{36}{193}{1940}

The DZP+diff/BLYP energy profiles for all three rearrangements are
shown in Fig.~4. A maximum step size of $0.1\,a_0$ was used for the left-hand side of the
acceptor-tunneling path. This value was increased to $0.15\,a_0$ for all the other paths,
resulting in slightly less smooth profiles but much faster execution time.
\titlea{5}{Water Hexamer}%
Liu \etal\ \cite1\ have recently identified a VRT band of \water{6}\ at $83\,$cm$^{-1}$
on the basis of an isotope mixture test.
\endnote{K.~Liu, M.~G.~Brown, C.~Carter, R.~J.~Saykally, J.~K.~Gregory and D.~C.~Clary}%
{Nature}{381}{501}{1996}\edref\hexamer%
In contrast to smaller water clusters the lowest energy isomer of the hexamer is probably
not cyclic, and the four lowest energy structures found by Tsai and Jordan \cite1\ lie
within an energy range of only $100\,$cm$^{-1}$.
\endnote{C.~J.~Tsai and K.~D.~Jordan}{\cpl}{213}{181-188}{1993}\edref\TandJ
The most accurate calculations conducted so far suggest that a `cage' structure lies lowest, followed
closely by `prism' and `book' forms (Fig.~5) \cite1.
\endnote{K.~Kim, K.~D.~Jordan and T.~S.~Zwier}{\jacs}{116}{11568-11569}{1994}\edref\KJZ
DMC calculations were used to find the vibrationally averaged rotational constants for each structure
using an empirical potential, and the best match was found for the cage structure. On this basis the
spectrum was assigned to the cage \rf\hexamer.
\begfig 4.0 cm
\figure{5}{Cage, prism and book forms of \water6.}%
\endfig

There are several unanswered questions concerning the interpretation of this experiment. First, if
a number of isomers lie so close together in energy, how is it that only one of them seems to be observed?
One possibility might be that the other isomers are also present in the beam, but do not have a spectral
feature in the range scanned experimentally. In fact, simulations reveal numerous isomers of all three morphologies
illustrated in Fig.~5 with different arrangements of the hydrogen bonds. The remaining
discussion will concentrate on the cage and the results of calculations using the ASP-W2 and TIP4P potentials
described in \S 3. The ASP-W2 form should be very similar to that employed in the DMC calculations of nuclear
dynamics by Liu \etal\ \rf\hexamer. However, we note that the cage isomer is also not the lowest in energy for
either of these potentials, although inclusion of zero-point energy can alter the ordering \rf\hexamer.
In fact, for the ASP-W2 potential the cyclic structure of the water pentamer is also not the global minimum,
but nevertheless, the lowest energy rearrangements of this isomer are quite well reproduced \rf\WWpent.
\begfig 9.4 cm
\figure{6}{Isomers of the cage structure for \water{6}\ calculated using the ASP-W2 potential with
binding energies in cm$^{-1}$.}%
\endfig

For the ASP-W2 potential only the first order induction energy was considered, since iterating this term
to convergence is time-consuming and was found to make no qualitative difference to results for the
water pentamer in previous work \rf\WWpent. There are then four isomers of the cage structure shown in Fig.~5
differing only in the position of the free hydrogen atoms of the two terminal, single-donor, single-acceptor
monomers (Fig.~6). No low energy degenerate rearrangements of these isomers were found, but transition
states were located for non-degenerate single flip and bifurcation mechanisms. For every single flip
process there is an analogous bifurcation which links different permutational isomers of the same structures.
Details of the paths are given in Table 3 and the single flip and bifurcation pathways linking cage isomers
C1 and C2 are illustrated in Fig.~7 and Fig.~8. The other paths are omitted for brevity.
\begfig 11.7 cm
\figure{7}{Single flip mechanism which interconverts C1 and C2 for the ASP-W2 potential.}%
\endfig
The experimental VRT transition has been assigned to a torsional motion of one of the 
single-donor, single-acceptor monomers, and the form of the spectrum has been described as
\lq near-prolate' \rf\hexamer. Note that even with full vibrational averaging over the four cage isomers
in Fig.~6 a symmetric top spectrum would not be expected.
Experimentally, every line was found to be split into a triplet with intensities in the ratio 9:6:1 and equal spacings 
of $1.92\,$MHz \rf\hexamer. Liu \etal\ have explained how this pattern might emerge in terms of
hypothetical degenerate rearrangements of the cage which they assumed must interchange the hydrogens
of two monomers in almost equivalent environments. This could lead to a doublet of doublets where the
middle lines are essentially superimposed. The resulting nuclear spin weights can then reproduce the
observed intensity pattern \rf\hexamer. 
\vskip 8 true mm \tabcap{3}{Rearrangement mechanisms which interconvert cage isomers of \water{6}\
calculated with the ASP-W2 potential.
The energies are in cm$^{-1}$. Min$_1$ is the lower minimum, $\Delta_1$
is the higher barrier, TS is the transition state and $\Delta_2$
is the smaller barrier corresponding to the higher minimum Min$_2$. $S$ is the integrated path
length in \AA, $D$ is the displacement between minima in \AA\ and $\gamma$ is the
cooperativity index. All these quantities are defined in \S 3.}
\centerline{
$$\vbox{\petit
\halign{
  \hfil    # \ \hfil&
  \hfil\  # \ \hfil&
  \hfil\  # \ \hfil&
  \hfil\  # \ \hfil&
  \hfil\  # \ \hfil&
  \hfil\ $#$\ \hfil&
  \hfil\ $#$\ \hfil&
  \hfil\ $#$\ \hfil&
       \quad  # \hfil \cr
   \noalign{\medskip}
   \noalign{\hrule}
   \noalign{\vskip1.5true pt}
   \noalign{\hrule}
   \noalign{\medskip}
    Min$_1$ & $\Delta_1$ & TS & $\Delta_2$ & Min$_2$ & S & D & \gamma & \hfil description \cr
   \noalign{\medskip}
   \noalign{\hrule}
   \noalign{\vskip1.5true pt}
   \noalign{\hrule}
   \noalign{\medskip}
       C1     &  560   &      -15,428   &   508  &   C2   &   2.8  &   2.1  &  13.7 &  single flip    \cr
       C1     &  641   &      -15,347   &   589  &   C2   &   2.9  &   2.2  &   7.6 &  bifurcation    \cr
       C1     &  323   &      -15,664   &   251  &   C3   &   1.7  &   1.3  &  15.4 &  single flip    \cr
       C1     &  1,135 &      -14,853   &   1,063&   C3   &   3.2  &   2.1  &   8.5 &  bifurcation    \cr
       C2     &  325   &      -15,612   &   285  &   C4   &   1.7  &   1.3  &  16.2 &  single flip    \cr
       C2     &  1,058 &      -14,878   &   1,018&   C4   &   3.4  &   2.1  &   8.6 &  bifurcation    \cr
       C3     &  524   &      -15,391   &   506  &   C4   &   2.7  &   2.0  &  14.1 &  single flip    \cr
       C3     &  650   &      -15,266   &   631  &   C4   &   2.7  &   2.2  &   7.9 &  bifurcation    \cr
  \noalign{\medskip\hrule\vskip1.5true pt\hrule}
 }  } $$
}
\vskip 8 true mm
\begfig 11.9 cm
\figure{8}{Bifurcation mechanism which interconverts C1 and C2 for the ASP-W2 potential.}%
\endfig

Liu \etal\ additionally suggested that the two monomers in question
might be the two terminal molecules which we have found to undergo flip and bifurcation rearrangements above.
However, no direct degenerate rearrangements of the cage isomers have been found to date. Since a combination
of sequential bifurcation  and flip rearrangements would achieve the desired effect we will now consider
the splittings that might result in more detail.
The eight single flip and bifurcation processes described above for the cage isomers link 16 versions:
four of each isomer. To distinguish between versions of each isomer we need only specify which of the
two hydrogens is free, and so we label the hydrogens on the terminal monomer
with two double-acceptor neighbours
a and b and those on the monomer with two double-donor neighbours c and d. The four relevant versions
of the C1 isomer may then be written C1(ac), C1(bc), C1(ad) and C1(bd). The interconnectivity of all
16 versions is shown in Fig.~9. If we make a H\"uckel-type approximation then the resulting
secular determinant is:
\vbox{\petit
$$ \pmatrix{ 0 & 0 & 0 & 0 & f_{12} & b_{12} & 0 & 0 & f_{13} & 0 & b_{13} & 0 & 0 & 0 & 0 & 0 \cr
             0 & 0 & 0 & 0 & b_{12} & f_{12} & 0 & 0 & 0 & f_{13} & 0 & b_{13} & 0 & 0 & 0 & 0 \cr
             0 & 0 & 0 & 0 & 0 & 0 & f_{12} & b_{12} & b_{13} & 0 & f_{13} & 0 & 0 & 0 & 0 & 0 \cr
             0 & 0 & 0 & 0 & 0 & 0 & b_{12} & f_{12} & 0 & b_{13} & 0 & f_{13} & 0 & 0 & 0 & 0 \cr
             f_{12} & b_{12} & 0 & 0 & 52 & 0 & 0 & 0 & 0 & 0 & 0 & 0 & f_{24} & 0 & b_{24} & 0 \cr
             b_{12} & f_{12} & 0 & 0 & 0 & 52 & 0 & 0 & 0 & 0 & 0 & 0 & 0 & f_{24} & 0 & b_{24} \cr
             0 & 0 & f_{12} & b_{12} & 0 & 0 & 52 & 0 & 0 & 0 & 0 & 0 & b_{24} & 0 & f_{24} & 0 \cr
             0 & 0 & b_{12} & f_{12} & 0 & 0 & 0 & 52 & 0 & 0 & 0 & 0 & 0 & b_{24} & 0 & f_{24} \cr
             f_{13} & 0 & b_{13} & 0 & 0 & 0 & 0 & 0 & 72 & 0 & 0 & 0 & f_{34} & b_{34} & 0 & 0 \cr
             0 & f_{13} & 0 & b_{13} & 0 & 0 & 0 & 0 & 0 & 72 & 0 & 0 & b_{34} & f_{34} & 0 & 0 \cr
             b_{13} & 0 & f_{13} & 0 & 0 & 0 & 0 & 0 & 0 & 0 & 72 & 0 & 0 & 0 & f_{34} & b_{34} \cr
             0 & b_{13} & 0 & f_{13} & 0 & 0 & 0 & 0 & 0 & 0 & 0 & 72 & 0 & 0 & b_{34} & f_{34} \cr
             f_{24} & 0 & b_{24} & 0 & f_{24} & 0 & b_{24} & 0 & f_{34} & b_{34} & 0 & 0 & 91 & 0 & 0 & 0 \cr
             0 & f_{24} & 0 & b_{24} & 0 & f_{24} & 0 & b_{24} & b_{34} & f_{34} & 0 & 0 & 0 & 91 & 0 & 0 \cr
             b_{24} & 0 & f_{24} & 0 & b_{24} & 0 & f_{24} & 0 & 0 & 0 & f_{34} & b_{34} & 0 & 0 & 91 & 0 \cr
             0 & b_{24} & 0 & f_{24} & 0 & b_{24} & 0 & f_{24} & 0 & 0 & b_{34} & f_{34} & 0 & 0 & 0 & 91 \cr } $$ }
where $f_{ij}$ and $b_{ij}$ are the appropriate tunneling matrix elements between C$i$ and C$j$.
We can simplify this problem by focusing on one of the isomers, say C1, and considering an effective
generator \cite1\ corresponding to a flip and a bifurcation (in either order). 
\endnote{B.~J.~Dalton and P.~D.~Nicholson}{\ijqc}{9}{325}{1975}
The reaction graph can then 
be represented as shown in Fig.~10, where the double lines indicate that there are two routes between
each pair of versions depending upon whether the flip or bifurcation comes first. 
The effective MS group contains the elements $E$, (ab), (bc) and (ab)(bc) and is isomorphic
to $C_{2v}$.  If the four connections 
are each represented by the same tunneling matrix element $\beta$ then the energy level pattern will
be identical to that obtained in the H\"uckel treatment of the $\pi$ system of butadiene, i.e.
$$ 2\beta (A_1), \qquad 0 (A_2,B_1), \qquad -2\beta (B_2), \eqno(5) $$
where the correspondence between the MS group elements and the operations of $C_{2v}$ has been
chosen as ${\rm (cd)}\equiv C_2$, ${\rm (ab)}\equiv \sigma_v(xz)$ and ${\rm (ab)(cd)}\equiv \sigma_v'(yz)$. 
The accidental degeneracy of the $A_2$ and $B_1$ states would be broken at higher resolution.
The relative
nuclear spin weights for rovibronic states are easily shown to be 9:3:3:1 for \water{6}\ and
4:2:2:1 for \Dwater{6} corresponding to
$A_1$:$A_2$:$B_1$:$B_2$. If the accidental degeneracy is unresolved then the relative intensities
of the three triplet components would be 9:6:1 for \water{6}\ and 4:4:1 for \Dwater{6}.
\begfig 11.0 cm
\figure{9}{Connectivity of versions of the cage isomers C1, C2, C3 and C4 according to the rearrangements
listed in Table 3. Solid lines represent the four single flip processes and dashed lines the four
bifurcations.}%
\endfig

The above analysis leads to the same result as that obtained by  Liu \etal\ \rf\hexamer, who considered the consequences
of hypothetical direct permutations (ab) and (cd). However, in the present picture the tunneling between
different versions of the four cage isomers does not occur through a single barrier, which 
might explain why the splittings are rather small. For this framework to be consistent we would expect to find
similar splitting patterns for all four cage isomers, although the magnitude of the splitting could be different
for each one. 

The same analysis would also hold for a second family of cage isomers where the double
acceptor monomers are adjacent. However, the four corresponding isomers in this case lie several hundred
wavenumbers higher in energy than those considered above. Several mechanisms interconverting the two cage families
were found in the set of 900 pathways calculated for the hexamer in this study. One further point worthy
of consideration is that the experiment measures the difference in tunneling splittings between different
rovibrational states. If the observed vibrational transition does indeed correspond to the torsional motion
of a single-donor, single-acceptor monomer then the splitting in the excited state could be significantly
different from the ground state.
\begfig 4.5 cm
\figure{10}{Effective reaction graph for the four versions of cage isomer C1 that are interconnected
indirectly by flip and bifurcation rearrangements.}%
\endfig

For the TIP4P potential the situation is slightly different because there are only two cage isomers rather
than four. The single-donor, single-acceptor that has two double-donor neighbours exhibits only one torsional
minimum intermediate between the two states found in C1-C4 above. The other single-donor, single-acceptor 
has the same two torsional states as in C1-C4, and undergoes single flip and bifurcation rearrangements
as before. For the single-donor, single-acceptor monomer in the intermediate torsional state a direct
degenerate rearrangement corresponding to a bifurcated transition state was found. The pathways are summarized in Table
4 where we label the two isomers C1$'$ and C2$'$. Illustrations of these rearrangements are omitted for brevity.
A degenerate bifurcation rearrangement of the C2$'$ isomer could not be located, despite starting a number
of transition state searches from points around the expected geometry.
\vskip 8 true mm \tabcap{4}{Rearrangement mechanisms which interconvert cage isomers of \water{6}\
calculated with the TIP4P potential. 
The energies are in cm$^{-1}$. Min$_1$ is the lower minimum, $\Delta_1$
is the higher barrier, TS is the transition state, and $\Delta_2$
is the smaller barrier corresponding to the higher minimum Min$_2$. $S$ is the integrated path
length in \AA, $D$ is the displacement between minima in \AA\ and $\gamma$ is the
cooperativity index. All these quantities are defined in \S 3.}
\centerline{
$$\vbox{\petit
\halign{
  \hfil   # \ \hfil&
  \hfil\  # \ \hfil&
  \hfil\  # \ \hfil&
  \hfil\  # \ \hfil&
  \hfil\  # \ \hfil&
  \hfil  $#$\ \hfil&
  \hfil\ $#$\ \hfil&
  \hfil\ $#$\ \hfil&
       \  #   \hfil \cr
   \noalign{\medskip}
   \noalign{\hrule}
   \noalign{\vskip1.5true pt}
   \noalign{\hrule}
   \noalign{\medskip}
    Min$_1$ & $\Delta_1$ & TS & $\Delta_2$ & Min$_2$ & S & D & \gamma & \hfil description \cr
   \noalign{\medskip}
   \noalign{\hrule}
   \noalign{\vskip1.5true pt}
   \noalign{\hrule}
   \noalign{\medskip}
       C1$'$ (-16,533)    &  14     &      -16,519    &   2    &   C2$'$ (-16,521)  &   1.3  &   1.2  &  14.1 &  single flip    \cr
       C1$'$ (-16,533)    &  825    &      -15,708    &   813  &   C2$'$ (-16,521)  &   3.6  &   2.2  &   8.6 &  bifurcation    \cr
       C1$'$ (-16,533)    &  1,361  &      -15,172    &   1,361&   C1$'$ (-16,533)  &   4.5  &   2.2  &   8.7 &  bifurcation    \cr
  \noalign{\medskip\hrule\vskip1.5true pt\hrule}
 }  } $$
}
\vskip 8 true mm
Although the topology of the PES is different for the TIP4P potential the above rearrangements could produce
the same tunneling splittings and intensity pattern as before, since there are still pathways linking all four versions
of each isomer. Unfortunately there is insufficient experimental information to distinguish the two possibilities,
and it is quite possible that neither scenario is correct. Due to the relatively large number of degrees of freedom
and the uncertainties associated with the choice of empirical potential more accurate theoretical treatments of
the dynamics will not be easy. However, it may not be a bad approximation to neglect relaxations of the rest of the
cage in considering the dynamics of the two single-donor, single-acceptor monomers. Two-dimensional quantum
calculations of the torsional dynamics of these two molecules were performed under this approximation using the C1 geometry
and the discrete variable representation \cite1\ (DVR).
\endnote{Z.~Ba\v ci\'c and J.~C.~Light}{Annu.~Rev.~Phys.~Chem.}{40}{469}{1989}
These calculations follow the three-dimensional DVR calculations of Sabo \etal\ \rf\Bacic\ for \water{3}\ 
in employing a model Hamiltonian in which only rotation about the hydrogen-bonded O${-}$H bond is permitted: 
$$ \hat{\cal H} = -B_{\rm eff}\left( {\partial^2\over\partial\phi_1^2} + {\partial^2\over\partial\phi_2^2} \right)
                   + V(\phi_1,\phi_2), \eqno(6) $$
where $\phi_1$ and $\phi_2$ are the two torsional angles in question and the value of $B_{\rm eff}$ was taken to be
$19.63\,$cm$^{-1}$ for H$_2$O and $9.82\,$cm$^{-1}$ for D$_2$O.
For each torsional degree of freedom the basis functions were chosen as:
$$ \psi(\phi) = {1\over\sqrt{2\pi}} e^{im\phi}, \qquad m=0,\ \pm1,\ldots,\ \pm N,\eqno(7)  $$
giving a total of $(2N+1)^2$ basis functions for a given value of $N$. The DVR grid points are then
uniformly spaced in each coordinate:
$$ \phi^j = { 2\pi j \over 2N+1}, \qquad j=1,\ 2,\ldots,\ 2N+1. \eqno(8) $$
Hence we obtain a direct product DVR with grid points $(\phi^j_1,\phi^k_2)$, and using the known
analytic form for the kinetic energy operator \cite1\ we obtain the Hamiltonian matrix elements:
$$ H^{ab}_{cd} = T_{ca}\delta_{db} + T_{db}\delta_{ca} + V(\phi_1^c,\phi_2^d)\delta_{ac}\delta_{db},\eqno(9)  $$
\endnote{D.~T.~Colbert and W.~H.~Miller}{\jcp}{96}{1982}{1992}
where 
$$ T_{ca} = B_{\rm eff} (-1)^{c-a} \cases{ N(N+1)/3, & $a=c$, \cr
                                           \noalign{\medskip}
                                           { \displaystyle\cos\left[\pi(c-a)/(2N+1)\right] \over 
                                             \displaystyle 2\sin^2\left[\pi(c-a)/(2N+1)\right] }, 
                                           & $a\not=c$. \cr } \eqno(10) $$
The lowest eigenvalues were converged to an accuracy better than $0.1\,$cm$^{-1}$ for both \water{6}\ 
and \Dwater{6}\ at a value $N=14$; an iterative Lanczos matrix diagonalization procedure was employed.
\begfig 17.7 cm
\figure{11}{Amplitudes of the lowest eight eigenvectors in torsional space found by two-dimensional DVR calculations
for \water{6}\ with the ASP-W2 potential. The interval between grid lines is $12.41^\circ$ in both directions.}%
\endfig

The lowest eight eigenvectors for \water{6}\ along with assignments are shown in Fig.~11 for ASP-W2 and
Fig.~12 for TIP4P. In both cases the ranges of the torsional angles have been restricted to exclude regions
where the amplitude is essentially always zero. The centre of each surface corresponds to the
geometry with both free hydrogens in intermediate positions. Not surprisingly, the wavefunctions for the
TIP4P potential are all delocalized over torsional space, and can be classified according to the number of
nodes in the $\phi_1$ and $\phi_2$ directions. The three lowest energy wavefunctions for the ASP-W2 potential
are localized in the wells corresponding to C1, C2 and C3. However, the fourth, seventh and eighth functions
appear to be delocalized over two isomers in each case. For this potential the wavefunctions are described
in terms of localized functions in the four different wells, e.g.~C1(1,0) is the function localized in 
the C1 well with one node in the $\phi_1$ direction and none in the $\phi_2$ direction.
The results for \Dwater{2}\ are omitted for brevity, and can be obtained from the author on request.
\begfig 17.7 cm
\figure{12}{Amplitudes of the lowest eight eigenvectors in torsional space found by two-dimensional DVR calculations
for \water{6}\ with the TIP4P potential. The interval between grid lines is $12.41^\circ$ in both directions.}%
\endfig

Clearly we cannot assign the experimental transition on the basis of these model calculations.
However, the fact that both potentials exhibit some delocalization between different cage isomers 
suggests that vibrational averaging of the structure is likely, especially in an excited torsional
state. Unfortunately, it does not seem to be possible to admit both the flip and bifurcation rearrangements
without including at least three angular degrees of freedom for each of the two monomers in question.
Such calculations are considerably more difficult, and are left for future work.

\begref{References}{80.}%
\references
\endref
\bigskip
\noindent Since this paper was written in October 1996 a number of additional
publications have appeared on the subject of small water clusters. 
The most relevant are collected here for convenience. 
\bigskip

\item{(a)} J. K. Gregory, \cpl\ {\bf 282}, 147 (1998).
\item{(b)} C. S. Guiang and R. E. Wyatt, \ijqc\ {\bf 68}, 233 (1998).
\item{(c)} I. M. Quintana, W. Ortiz and G. E. Lopez, \cpl\ {\bf 287}, 429 (1998).
\item{(d)} J. B. Paul, R. A. Provencal and R. J. Saykally, \jpca\ {\bf 102}, 3279 (1998).
\item{(e)} O. Engkvist, N. Forsberg, M. Sch\"utz and G. Karlstrom, \molphys\ {\bf 90}, 277 (1997).
\item{(f)} J. K. Gregory and D. C. Clary, \jpca\ {\bf 101}, 6813 (1997).
\item{(g)} J. K. Gregory, D. C. Clary, K. Liu, M. G. Brown and R. J. Saykally, Science {\bf 275}, 814 (1997).
\item{(h)} M. P. Hodges, A. J. Stone and S. S. Xantheas, \jpca\ {\bf 101}, 9163 (1997).
\item{(i)} C. Leforestier, L. B. Braly, K. Liu, M. J. Elrod and R. J. Saykally, \jcp\ {\bf 106}, 8527 (1997).
\item{(j)} M. Sch\"utz, S. Brdarski, P.-O. Widmark, R. Lindh and G. Karlstr\"om, \jcp\ {\bf 107}, 4597 (1997).
\item{(k)} J. B. Paul,  C. P. Collier,  R. J. Saykally, J. J. Scherer and  A. O'Keefe, \jpca\ {\bf 101}, 5211 (1997).
\item{(l)} M. R. Viant, J. D. Cruzan, D. D. Lucas, M. G. Brown, K. Liu and R. J. Saykally, \jpca\ {\bf 101}, 9032 (1997).
\item{(m)} J. D. Cruzan, M. R. Viant, M. G. Brown and R. J. Saykally, \jpca\ {\bf 101}, 9022 (1997).
\item{(n)} K. Liu, M. G. Brown and R. J. Saykally, \jpca\ {\bf 101}, 8995 (1997).
\item{(o)} J.~D.~Cruzan,  L.~B.~Braly, K.~Liu, M.~G.~Brown, J.~G.~Loeser and R.~J.~Saykally, Science {\bf 271}, 59 (1996).
\item{(p)} E.~Fredj, R.~B.~Gerber and M.~A.~Ratner, \jcp\ {\bf 105}, 1121 (1996).
\item{(q)} J.~K.~Gregory and  D.~C.~Clary,  \jpc\ {\bf 100}, 18014 (1996).
\item{(r)} K.~Liu, M.~G.~Brown, J.~D.~Cruzan and R.~J.~Saykally, Science {\bf 271}, 62 (1996).
\item{(s)} K.~Liu, M.~G.~Brown, M.~R.~Viantm, J.~D.~Cruzan and R.~J.~Saykally, Mol.~Phys.~{\bf 89}, 1373 (1996).
\item{(t)} E.~M.~Mas and K.~Szalewicz, \jcp\ {\bf 104}, 7606 (1996).
\item{(u)} E.~H.~T.~Olthof, A.~Van der Avoird, P.~E.~S.~Wormer, K.~Liu and R.~J.~Saykally, \jcp\ {\bf 105}, 8051 (1996).
\item{(v)} J.~M.~Pedulla, F.~Vila and K.~D.~Jordan, \jcp\ {\bf 105}, 11091 (1996).
\item{(w)} D. Sabo, Z.~Ba\v ci\'c, T.~B\"urgi and S.~Leutwyler, \cpl\ {\bf 261}, 318 (1996).
\item{(x)} J.~M.~Sorenson, J.~K.~Gregory and D.~C.~Clary, \cpl\ {\bf 263}, 680 (1996).
\item{(y)} A.~Van der Avoird, E.~H.~T.~Olthof and P.~E.~S.~Wormer, \jcp\ {\bf 105}, 8034 (1996).

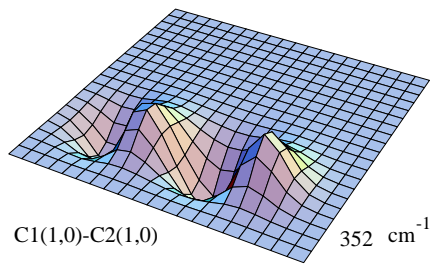
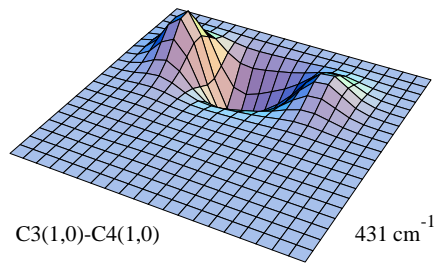
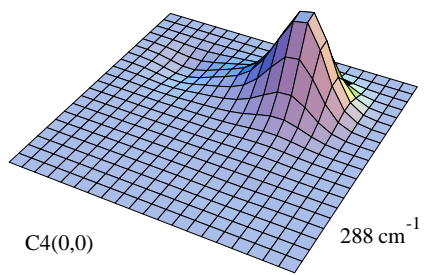
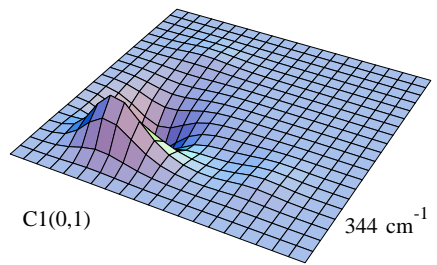
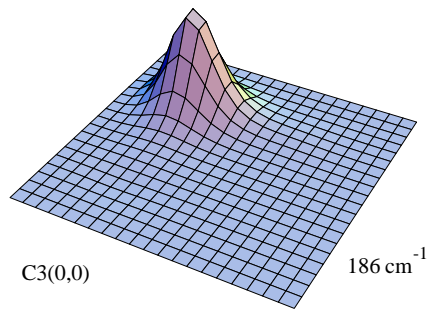
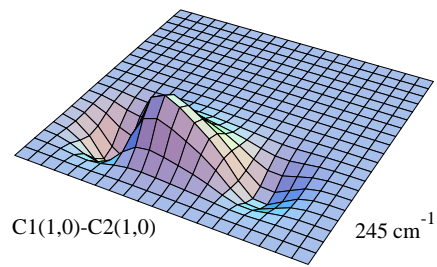
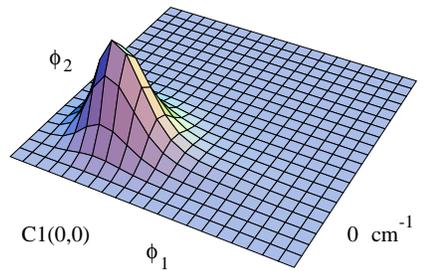
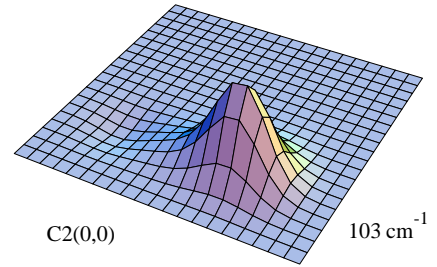
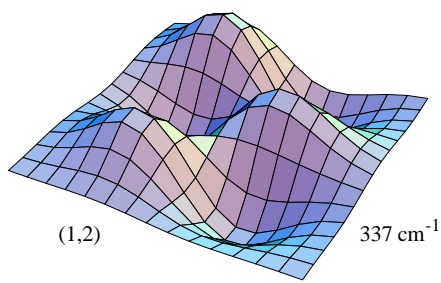
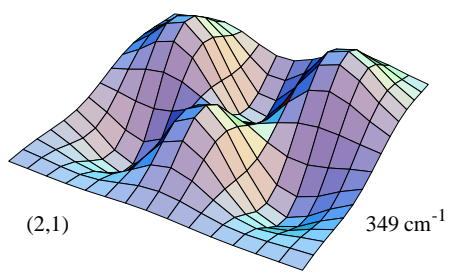
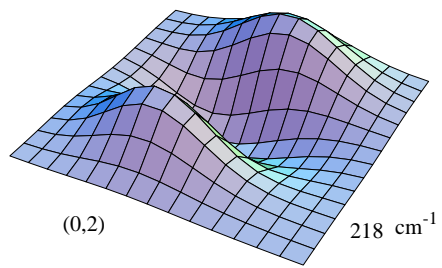
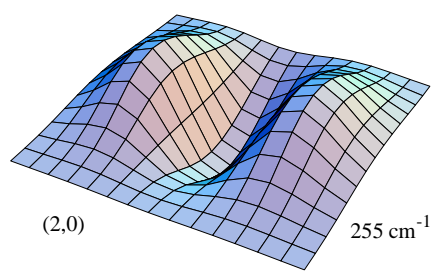
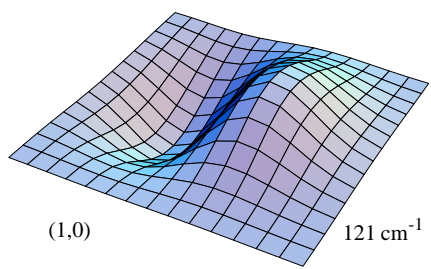
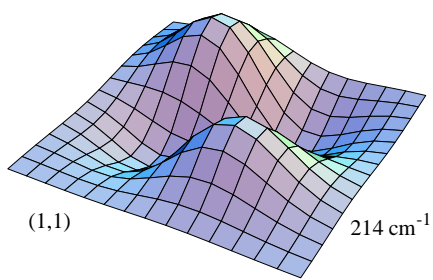
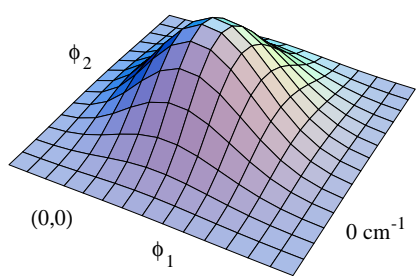
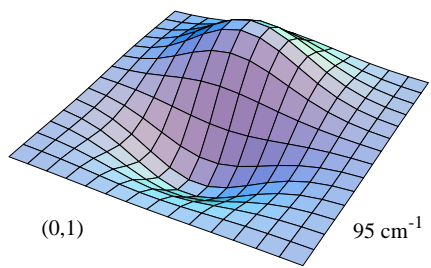
\bye